\documentstyle[12pt]{article}

\begin{document}

\input epsf

\newcommand{\beq}{\begin{equation}}
\newcommand{\eeq}{\end{equation}}
\newcommand{\bea}{\begin{eqnarray}}
\newcommand{\eea}{\end{eqnarray}}
\newcommand{\eL}{{\cal L}}
\newcommand{\half}{\textstyle{1 \over 2}}
\newcommand{\J}{\bf J}
\newcommand{\bP}{\bf P}
\newcommand{\G}{\bf G}
\newcommand{\K}{\bf K}
\newcommand{\M}{{\cal M}}
\newcommand{\bu}{\bf u}
\newcommand{\la}{\lambda}

\begin{flushright}
physics/9708015 \\
DOE-ER-40757-104\\
UTEXAS-HEP-97-18
\end{flushright}

\begin{center}
{\Large {\bf The Geometry of $SU(3)$}}
\end{center}

\hfil\break

\begin{center}
{\bf Mark Byrd\footnote{mbyrd@physics.utexas.edu}\\}
\hfil\break
{\it Center for Particle Physics \\
University of Texas at Austin \\
Austin, Texas 78712-1081}
\end{center}

\begin{abstract}

The group $SU(3)$ is parameterized in terms of generalized ``Euler angles''.  The differential operators of $SU(3)$ corresponding to the 
Lie Algebra elements are obtained, the invariant forms are found,
the group invariant volume element is found,
 and some relevant comments about the geometry of the group manifold are made.

\end{abstract}

\section{Introduction}

For a long time the algebra of $SU(3)$ has been studied for its application to physics.  It's important to particle physics not only due to the now famous Eight-fold way developed by Gell-Mann and Ne'eman \cite{GN}, which has become known as flavor $SU(3)$, but also due to the color $SU(3)$ of QCD (\cite{W} contains an excellent review along with a meticulous set of references).  Color $SU(3)$ was introduced for consistency in the quark model to give the quark wave functions their observed fermionic nature within the flavor framework.  The color $SU(3)$ is believed to be an exact gauge symmetry, whereas the flavor symmetry is broken by the quark masses.  They are both very important to our present understanding of particle physics as well as is the $SU(3)$ nuclear physics model.  This is based on the $SU(3)$ symmetry group of the 3-d isotropic harmonic oscillator.
  Until now the algebra of $SU(3)$ has been studied extensively but the geometry of the group manifold has had much less attention.  Hopefully that will eventually change.  The work done here could help to further the use of the group itself.  The explicit forms of the left and right invariant vector fields and left and right invariant one-forms are given in terms of the Euler angle parameterization.  These should help with any subsequent study of the group manifold.  

To summarize, in section 2 the method for determining the left and right invariant vector fields of $SU(3)$ is discussed and the explicit forms of the operators are given in terms of the Euler angle parameterization.  In section 3 the group invariant forms are given enabling the derivation of the group invariant volume element, i.e., the Haar measure in section 4.  Finally, in section 5, some of the many applications are discussed.

\section{The Lie Algebra of $SU(3)$}

The Lie Algebra of a group is a set of left invariant vector fields on the group manifold.  These can be constructed by a method that has been applied to $SU(2)$ (see for example \cite{lb}) and can, in principle, be used for $SU(n)$.  Here it is done explicitly for $SU(3)$.  

\subsection{``Euler Angle'' Decomposition}

To decompose the group into the ``Euler angle'' parameterization we first consider the algebra.  The algebra must obey the commutation relations given in the table below.  The Gell-Mann matrices provide the most common representation in terms of $3\times3$ hermitian, traceless matrices.  This set is linearly independent and is the lowest dimensional faithful representation of the algebra.  These are:
$$
\begin{array}{crcr}

\lambda_1 = \left( \begin{array}{crcl}
                     0 & 1 & 0 \\
                     1 & 0 & 0 \\
                     0 & 0 & 0   \end{array} \right), &

\lambda_2 = \left( \begin{array}{crcr} 
                     0 & -i & 0 \\
                     i &  0 & 0 \\
                     0 &  0 & 0   \end{array} \right), &

\lambda_3 =  \left( \begin{array}{crcr} 
                     1 &  0 & 0 \\
                     0 & -1 & 0 \\
                     0 &  0 & 0   \end{array} \right), \\

\lambda_4 =  \left( \begin{array}{clcr} 
                     0 & 0 & 1 \\
                     0 & 0 & 0 \\
                     1 & 0 & 0   \end{array} \right), &

\lambda_5 = \left( \begin{array}{crcr} 
                     0 & 0 & -i \\
                     0 & 0 & 0 \\
                     i & 0 & 0   \end{array} \right), &

 \lambda_6 = \left( \begin{array}{crcr} 
                     0 & 0 & 0 \\
                     0 & 0 & 1 \\
                     0 & 1 & 0   \end{array} \right), \\

\lambda_7 = \left( \begin{array}{crcr} 
                     0 & 0 & 0 \\
                     0 & 0 & -i \\
                     0 & i & 0   \end{array} \right), &

\lambda_8 = \frac{1}{\sqrt{3}}\left( \begin{array}{crcr} 
                     1 & 0 & 0 \\
                     0 & 1 & 0 \\
                     0 & 0 & -2   \end{array} \right).

\end{array}    
$$
From these, one may then work out the commutation relations that are generally valid for any representation of the algebra.
$$
\hfil\break
\begin{array}{|c||c|c|c|c|c|c|c|c|} \hline
         & \la_1    &    \la_2 &    \la_3 &    \la_4 &    \la_5 &    \la_6 &    \la_7 &    \la_8     \\ \hline \hline
\la_1    &     0    &  2i\la_3 & -2i\la_2 &   i\la_7 &  -i\la_6 &   i\la_5 &  -i\la_4 &            0 \\ 
         &          &          &          &          &          &          &          &              \\ \hline
\la_2    & -2i\la_3 &        0 &  2i\la_1 &   i\la_6 &   i\la_7 &  -i\la_4 &  -i\la_5 &            0 \\ 
         &          &          &          &          &          &          &          &              \\ \hline
\la_3    &  2i\la_2 & -2i\la_1 &        0 &   i\la_5 &  -i\la_4 &  -i\la_7 &   i\la_6 &            0 \\ 
         &          &          &          &          &          &          &          &              \\ \hline
\la_4    &  -i\la_7 &  -i\la_6 &  -i\la_5 &        0 &   i\la_3 &   i\la_2 &   i\la_1 & -i\sqrt{3}\la_5 \\
         &          &          &          &          &+i\sqrt{3}\la_8&     &          &              \\ \hline
\la_5    &  i\la_6  &  -i\la_7 &   i\la_4 &  -i\la_3 &       0  &  -i\la_1 &  -i\la_2 & i\sqrt{3}\la_4 \\
         &          &          &          &-i\sqrt{3}\la_8&     &          &          &               \\ \hline
\la_6    & -i\la_5  &   i\la_4 &   i\la_7 &  -i\la_2 &   i\la_1 &        0 &  -i\la_3 &-i\sqrt{3}\la_7 \\ 
         &          &          &          &          &          &          &+i\sqrt{3}\la_8&           \\ \hline
\la_7    &  i\la_4   &   i\la_5 &  -i\la_6 &  -i\la_1 &   i\la_2 &   i\la_3 &        0 & i\sqrt{3}\la_6 \\ 
         &          &          &          &          &          &-i\sqrt{3}\la_8&     &               \\ \hline
\la_8    &      0   &        0 &        0 &i\sqrt{3}\la_5&-i\sqrt{3}\la_4&i\sqrt{3}\la_7&-i\sqrt{3}\la_6&0 \\ 
         &          &          &          &          &          &          &          &              \\ \hline
\end{array}
$$
\hfil\break
The entries in the table are given by commuting the element in the first column, (i.e., the element of the algebra in the column separated by the double line) with the element in the top row (which is also separated by a double line). 
The reason for displaying the whole table is two-fold.  First, it is easy to 
read off the structure constants, defined by:
$$
\left[ \la_i, \la_j \right] = C^k_{\;ij} \la_k.
$$
Second, we can see the relationship in the algebra that defines a so-called Cartan decomposition.  That is, for subsets of the group manifold, $K$, and $P$, there corresponds $\eL(K)$ and $\eL(P)$, subsets of the Lie Algebra of $SU(3)$, denoted here $\eL(SU(3))$, such that for $k_1, k_2 \in \eL(K)$, and $p_1, p_2 \in \eL(P)$,
$$
\left[ k_1, k_2 \right] \in \eL(K), \;\;\;\;\;\;\;\left[ p_1, p_2 \right] \in \eL(K),
$$
and
$$
\left[ k_1, p_1 \right] \in \eL(P).
$$
Here, $\eL(K) = \{ \la_1, \la_2, \la_3, \la_8\}$, and $\eL(P) = \{\la_4, \la_5, \la_6, \la_7 \}$.  Given the decomposition of the algebra into a (semi-) direct sum,
$$
\eL(G) = \eL(K) \oplus \eL(P),
$$
we have a decomposition of the group,
$$
G = K \cdot P.
$$ 
Here
$K$ is the $SU(2)$ subgroup obtained by exponentiating the corresponding algebra \{$\la_1,\la_2,\la_3$\} plus $\la_8$.  Since $\la_8$ commutes with this $SU(2)$ subgroup, this can be written as
$$
K= e^{(i\la_3 \alpha)} e^{(i\la_2 \beta)}e^{(i\la_3 \gamma)} e^{(i\la_8 \phi)}.
$$
Then we may decompose $P$.  Of the four elements of $P$, we can pick one, say $\la_5$, analogous to the $J_y =J_2$ of $SU(2)$ and write any element of $P$ as 
$$
P=K^{\prime} e^{(i\la_5 \theta)}K^{\prime},
$$
where $K^{\prime}$ is another copy of $K$.  Dropping the redundancies,
 we arrive at the following product for $SU(3)$. (This proof was adapted from the book by Hermann \cite{RH} and first pointed out to me by Biedenharn \cite{lb2}.)  
\beq
D(\alpha,\beta,\gamma,\theta,a,b,c,\phi) = e^{(i\la_3 \alpha)} e^{(i\la_2 \beta)}e^{(i\la_3 \gamma)} e^{(i\la_5 \theta)} e^{(i\la_3 a)} e^{(i\la_2 b)} 
e^{(i\la_3 c)} e^{(i\la_8 \phi)},
\eeq
where $D$ is an arbitrary element of $SU(3)$.  This can be written as
$$
D^{(3)}(\alpha,\beta,\gamma,\theta,a,b,c,\phi) = D^{(2)}(\alpha,\beta,\gamma)
e^{(i\la_5 \theta)}D^{(2)}(a,b,c)e^{(i\la_8 \phi)},
$$
where the $D^{(3)}$ denotes an arbitrary element of $SU(3)$ and $D^{(2)}$ is an arbitrary element of $SU(2)$ as a subset of $SU(3)$.

As stated above, this method could, in principle, be used to decompose $SU(n)$ into an Euler angle type parameterization.  This would enable the calculation of the left and right invariant vector fields and one forms as discussed below.

\subsection{The Vector Fields}

The way to find the differential operators corresponding to the Lie algebra 
elements is now rather simple but messy.  To do this, take derivatives of 
$D(\alpha,\beta,\gamma,\theta,a,b,c,\phi)$ in (1) with respect to each of its parameters.
For brevity, I'll call this just $D$.  One then obtains the following results, using the Baker-Campbell-Hausdorff relation,:
\bea
\frac{\partial D}{\partial \alpha} = -i\la_3 D,
\eea
\bea
\frac{\partial D}{\partial \beta}
       &=&e^{(-i\la_3 \alpha)} (-i\la_2) e^{(i\la_3 \alpha)} D \nonumber \\
       &=&-i(-\la_1 \sin{2\alpha} + \la_2 \cos{2\alpha})D,
\eea
and
\bea
\frac{\partial D}{\partial \gamma} &=& e^{(-i\la_3 \alpha)}e^{(-i\la_2 \beta)}
(-i\la_3)e^{(i\la_2 \beta)}e^{(i\la_3 \alpha)}D\nonumber\\
&=& -i(\la_1 \cos{2\alpha} \sin{2\beta} + \la_2 \sin{2\alpha} \sin{\beta}
 +\la_3 \cos{2\beta})D.
\eea
These are the first three.  Continuing this, we obtain linear combinations of the 
Lie algebra elements, (the $\la$'s).  The differential operators are then given by
\beq
 \Lambda_i D = -\la_i D 
\eeq
Where the minus sign is on the $\la$ acting as a differential operator\footnote{There is an excellent discussion of this procedure for the $SU(2)$ case in \cite{lb}.  The methods used here follow that treatment closely.}.  The eight equations and eight unknowns may then be solved to find the following vector fields(differential operators)corresponding to the Lie algebra elements.  With  $\partial_1 \equiv \frac{\partial}{\partial \alpha}$, etc.,
\bea
\Lambda_1&=&i \cos2\alpha \cot2\beta \partial_1 + i \sin2\alpha \partial_2 - 
i\frac{\cos2\alpha}{\sin2\beta} \partial_3 \\ 
\Lambda_2&=&-i \sin2\alpha \cot2\beta \partial_1+i\cos2\alpha\partial_2+i
\frac{\sin2\alpha}{\sin2\beta}\partial_3 \\
\Lambda_3&=&i\partial_1
\eea
\bea
\Lambda_4&=&i\frac{\sin\beta}{\sin2\beta}\cot\theta \cos(\alpha+\gamma) \partial_1
-i\sin\beta \cot\theta \sin(\alpha+\gamma)\partial_2 \nonumber \\
         &-&i\cot2\beta \sin\beta \cot\theta \cos(\alpha+\gamma)\partial_3
+i\frac{(2-\sin^2\theta)}{\sin2\theta}\cos\beta \cos(\alpha+\gamma)\partial_3\nonumber \\
	 &+&i\cos\beta \sin(\alpha+\gamma)\partial_4\nonumber \\
         &-&i2\frac{\cos\beta}{\sin2\theta} \cos(\alpha + \gamma)\partial_5
-i\frac{\cot2b}{\sin\theta}\sin\beta \cos(\alpha-\gamma-2a)\partial_5\nonumber \\
         &+&i\frac{\sin\beta}{\sin\theta}\sin(\alpha-\gamma-2a)\partial_6\nonumber \\
         &+&i\frac{\sin\beta}{\sin\theta \sin2b}\cos(\alpha-\gamma-2a)\partial_7\nonumber \\
         &-&\frac{\sqrt{3}}{2}\tan\theta \cos\beta \cos(\alpha+\gamma)\Lambda_8
\eea
\bea
\Lambda_5&=&-i\frac{\sin\beta}{\sin2\beta}\cot\theta \sin(\alpha+\gamma) \partial_1
            -i\sin\beta \cot\theta \cos(\alpha+\gamma)\partial_2 \nonumber \\
         &+&i\cot2\beta \sin\beta \cot\theta \sin(\alpha+\gamma)\partial_3
          -i\frac{(2-\sin^2\theta)}{\sin2\theta}\cos\beta \sin(\alpha+\gamma)\partial_3\nonumber\\
         &+&i\cos\beta \cos(\alpha+\gamma)\partial_4\nonumber \\
         &+&i2\frac{\cos\beta}{\sin2\theta} \sin(\alpha + \gamma)\partial_5
          +i\frac{\cot2b}{\sin\theta}\sin\beta \sin(\alpha-\gamma-2a)\partial_5\nonumber \\
         &+&i\frac{\sin\beta}{\sin\theta}\cos(\alpha-\gamma-2a)\partial_6\nonumber \\
         &-&i\frac{\sin\beta}{\sin\theta \sin2b}\sin(\alpha-\gamma-2a)\partial_7\nonumber \\
         &+&\frac{\sqrt{3}}{2}\tan\theta \cos\beta \sin(\alpha+\gamma)\Lambda_8
\eea
\bea
\Lambda_6&=&i\frac{\cos\beta}{\sin2\beta}\cot\theta \cos(\alpha-\gamma) \partial_1
            +i\cos\beta \cot\theta \sin(\alpha-\gamma)\partial_2 \nonumber \\
         &-&i\cot2\beta \cos\beta \cot\theta \cos(\alpha-\gamma)\partial_3
          -i\frac{(2-\sin^2\theta)}{\sin2\theta}\sin\beta \cos(\alpha-\gamma)\partial_3\nonumber \\
	 &+&i\sin\beta \sin(\alpha-\gamma)\partial_4\nonumber \\
         &+&i2\frac{\sin\beta}{\sin2\theta} \cos(\alpha - \gamma)\partial_5
          -i\frac{\cot2b}{\sin\theta}\cos\beta \cos(\alpha+\gamma+2a)\partial_5\nonumber \\
         &-&i\frac{\cos\beta}{\sin\theta}\sin(\alpha+\gamma+2a)\partial_6\nonumber \\
         &+&i\frac{\cos\beta}{\sin\theta \sin2b}\cos(\alpha+\gamma+2a)\partial_7\nonumber \\
         &+&\frac{\sqrt{3}}{2}\tan\theta \sin\beta \cos(\alpha-\gamma)\Lambda_8
\eea
\bea
\Lambda_7&=&i\frac{\cos\beta}{\sin2\beta}\cot\theta \sin(\alpha-\gamma) \partial_1
           -i\cos\beta \cot\theta \cos(\alpha-\gamma)\partial_2 \nonumber \\
         &-&i\cot2\beta \cos\beta \cot\theta \sin(\alpha-\gamma)\partial_3
          -i\frac{(2-\sin^2\theta)}{\sin2\theta}\sin\beta \sin(\alpha-\gamma)\partial_3\nonumber \\
         &-&i\sin\beta \cos(\alpha-\gamma)\partial_4\nonumber \\
         &+&i2\frac{\sin\beta}{\sin2\theta} \sin(\alpha - \gamma)\partial_5
          -i\frac{\cot2b}{\sin\theta}\cos\beta \sin(\alpha+\gamma+2a)\partial_5\nonumber \\
         &+&i\frac{\cos\beta}{\sin\theta}\cos(\alpha+\gamma+2a)\partial_6\nonumber \\
         &+&i\frac{\cos\beta}{\sin\theta \sin2b}\sin(\alpha+\gamma+2a)\partial_7\nonumber \\
         &+&\frac{\sqrt{3}}{2}\tan\theta \sin\beta \sin(\alpha-\gamma)\Lambda_8
\\
\Lambda_8 &=&i\sqrt{3}\partial_3 - i\sqrt{3}\partial_5 + i\partial_8\\
          & &               \nonumber
\eea
The right differential operators (the differential operators that correspond to
 this action when acting {\it from} the right) are different and are denoted 
$\Lambda_i^r$.
One may find these ``right'' differential operators in two ways.  First, one may use the relation 
$$
D\lambda_i = -\Lambda_i^rD,
$$
and do the same calculation as with the left.  Second, one may use the fact that
\beq
\Lambda_i^r = R_{ij}\Lambda_j,
\eeq
where $R_{ij}\in SO(8)$ is an element of the adjoint representation of $SU(3)$.  It is therefore a function of the eight parameters above.  The right invariant vector fields are then given by the following equations.
\bea
\Lambda^r_1&=&-i\cos 2c \cot 2b \partial_7 -i\sin 2c \partial_6 +i 
\frac{\cos 2c}{\sin 2b}\partial_5 \\
\Lambda_2^r&=&-i\sin 2c \cot 2b \partial_7 + i\cos 2c \partial_6+i
\frac{\sin 2c}{\sin 2b}\partial_5 \\
\Lambda_3^r&=&i\partial_7
\eea
\bea
\Lambda_4^r&=&-i\frac{\sin b}{\sin 2b}\cot \theta \cos(c+a+3\eta) \partial_7\nonumber\\
           &+&i\sin b \cot \theta \sin(c+a+3\eta) \partial_6\nonumber\\
           &+&i\cot 2b \sin b \cot \theta \cos(c+a+3\eta) \partial_5
            -i\frac{(2-\sin^2 \theta)}{\sin 2 \theta} \cos b \cos(c+a+3\eta)\partial_5\nonumber\\
           &-&i\cos b \sin(c+a+3\eta) \partial_4\nonumber\\
           &+&i2\frac{\cos b}{\sin 2\theta}\cos(c+a+3\eta)\partial_3
            +i\frac{\cot 2 \beta}{\sin \theta}\sin b\cos(c-a-2\gamma+3\eta)\partial_3\nonumber\\
           &-&i\frac{\sin b}{\sin \theta}\sin(c-a-2\gamma+3\eta)\partial_2\nonumber\\
           &-&i\frac{\sin b}{\sin \theta \sin 2\beta} \cos(c-a-2\gamma +3\eta)
\partial_1\nonumber\\
           &-&\frac{\sqrt{3}}{2}\tan \theta \cos b \cos(c+a+3\eta)\Lambda_8^r
\eea
\bea
\Lambda_5^r&=&-i\frac{\sin b}{\sin 2b}\cot \theta \sin(c+a+3\eta) \partial_7\nonumber\\
           &-&i\sin b \cot \theta \cos(c+a+3\eta) \partial_6\nonumber\\
           &+&i\cot 2b \sin b \cot \theta \sin(c+a+3\eta) \partial_5
            -i\frac{(2-\sin^2 \theta)}{\sin 2 \theta} \cos b \sin(c+a+3\eta)\partial_5\nonumber\\
           &+&i\cos b \cos(c+a+3\eta) \partial_4\nonumber\\
           &+&i2\frac{\cos b}{\sin 2\theta}\sin(c+a+3\eta)\partial_3
            +i\frac{\cot 2 \beta}{\sin \theta}\sin b\sin(c-a-2\gamma+3\eta)\partial_3\nonumber\\
           &+&i\frac{\sin b}{\sin \theta}\cos(c-a-2\gamma+3\eta)\partial_2\nonumber\\
           &-&i\frac{\sin b}{\sin \theta \sin 2\beta} \sin(c-a-2\gamma +3\eta)
\partial_1\nonumber\\
           &-&\frac{\sqrt{3}}{2}\tan \theta \cos b \sin(c+a+3\eta)\Lambda_8^r
\eea

\bea
\Lambda_6^r&=&i\frac{\cos b}{\sin 2b}\cot \theta \cos(c-a-3\eta) \partial_7\nonumber\\
           &+&i\cos b \cot \theta \sin(c-a-3\eta) \partial_6\nonumber\\
           &-&i\cot 2b \cos b \cot \theta \cos(c-a-3\eta) \partial_5
            -\frac{(2-\sin^2 \theta)}{\sin 2 \theta} \sin b \cos(c-a-3\eta)\partial_5\nonumber\\
           &+&i\sin b \sin(c-a-3\eta) \partial_4\nonumber\\
           &+&i2\frac{\sin b}{\sin 2\theta}\cos(c-a-3\eta)\partial_3
            -i\frac{\cot 2 \beta}{\sin \theta}\cos b\cos(c+a+2\gamma-3\eta)\partial_3\nonumber\\
           &-&i\frac{\cos b}{\sin \theta}\sin(c+a+2\gamma-3\eta)\partial_2\nonumber\\
           &+&i\frac{\cos b}{\sin \theta \sin 2\beta} \cos(c+a+2\gamma -3\eta)
\partial_1\nonumber\\
           &-&\frac{\sqrt{3}}{2}\tan \theta \sin b \cos(c-a-3\eta)\Lambda_8^r
\eea

\bea
\Lambda_7^r&=&-i\frac{\cos b}{\sin 2b}\cot \theta \sin(c-a-3\eta) \partial_7\nonumber\\
           &+&i\cos b \cot \theta \cos(c-a-3\eta) \partial_6\nonumber\\
           &+&i\cot 2b \cos b \cot \theta \sin(c-a-3\eta) \partial_5
            +i\frac{(2-\sin^2 \theta)}{\sin 2 \theta} \sin b \sin(c-a-3\eta)\partial_5\nonumber\\
           &+&i\sin b \cos(c-a-3\eta) \partial_4\nonumber\\
           &-&i2\frac{\sin b}{\sin 2\theta}\sin(c-a-3\eta)\partial_3
            +i\frac{\cot 2 \beta}{\sin \theta}\cos b\sin(c+a+2\gamma-3\eta)\partial_3\nonumber\\
           &-&i\frac{\cos b}{\sin \theta}\cos(c+a+2\gamma-3\eta)\partial_2\nonumber\\
           &-&i\frac{\cos b}{\sin \theta \sin 2\beta} \sin(c+a+2\gamma -3\eta)
\partial_1\nonumber\\
           &+&\frac{\sqrt{3}}{2}\tan \theta \sin b \sin(c-a-3\eta)\Lambda_8^r \\
\Lambda_8^r&=&i\partial_8
\eea
Here, $\eta = \phi/ \sqrt{3}$.  Note also that the right operators obey the commutation relation $[\Lambda^r_i,\Lambda^r_j]=-C^k_{ij}\Lambda^r_k$ (see \cite{lb} for a complete discussion).  

The calculation of the left invariant vector fields acting on the $D$ matrices was first attempted by T.J. Nelson \cite{N}.  However, the assumption of a singlet state for each irreducible representation was assumed.  This shortcomming is well noted by the author and was {\it not} used here.

\section{The Invariant Forms}

The left invariant forms for the manifold are dual to the left invariant vector fields.
Take the left invariant vector fields that make up the Lie algebra
$$
\lambda_i = a^j_i \partial_j
$$
and use the duality between the tangent and cotangent vectors to construct left invariant one forms.  If we take the left invariant one forms to have the form
\bea
\omega^{l} = b^l_k dx^k
\eea
then
\bea
\delta^l_i = < \omega_i,\lambda^{l} > = b^l_k a^j_i < d\alpha^j,\partial_k> = 
b^l_k a^j_i \delta_j^k,
\eea
therefore the matrices $b$ and $a$ are inverse transposes of each other.  In this way, we obtain left(and analogously right) invariant forms on the group manifold.  The left invariant forms are given by:

\bea
\omega^1&=&
\sin (2\,\alpha )d\beta - 
  \cos (2\,\alpha )\,\sin (2\,\beta )d\gamma \nonumber \\
&-& 
  \cos (2\,\alpha )\,\sin (2\,\beta )\,
    \left( 1 - \half{\sin^2 (\theta) } \right) da   \nonumber \\
&+& \cos (2\,a + 2\,\gamma )\,\cos (\theta )\,
      \sin (2\,\alpha ) db \nonumber \\
&+& \cos (2\,\alpha )\,\cos (2\,\beta )\,\cos (\theta )\,
      \sin (2\,a + 2\,\gamma ) db   \nonumber \\
&-& \left[ \cos (2\,\alpha )\,
        \cos (2\,\beta )\,\cos (2\,a + 2\,\gamma )\,
        \cos (\theta )\,\sin (2\,b) \right]dc \nonumber \\
&+& \cos (\theta )\,\sin (2\,\alpha )\,\sin (2\,b)\,
      \sin (2\,a + 2\,\gamma ) dc \nonumber \\
&-& \cos (2\,\alpha )\,\cos (2\,b)\,\sin (2\,\beta )\,
      \left( 1 - \half{\sin^2 (\theta) } \right) dc \nonumber \\
&+& \frac{\sqrt{3}}{2}\cos (2\,\alpha )\,
      \sin (2\,\beta )\,\sin^2 \theta d\phi \\
\omega^2&=&
\cos (2\,\alpha )d\beta + 
  \sin (2\,\alpha )\,\sin (2\,\beta )d\gamma \nonumber \\
&+&  \sin (2\,\alpha )\,\sin (2\,\beta )\,
   \left( 1 - \half\sin^2 (\theta ) \right)da  \nonumber \\
&+&  \cos (2\,\alpha )\,
      \cos (2\,a + 2\,\gamma )\,\cos (\theta ) db \nonumber \\
&-&  \cos (2\,\beta )\,\cos (\theta )\,\sin (2\,\alpha )\,
      \sin (2\,a + 2\,\gamma ) db  \nonumber \\
&+&  \cos (2\,\beta )\,\cos (2\,a + 2\,\gamma )\,
      \cos (\theta )\,\sin (2\,\alpha )\,\sin (2\,b) dc \nonumber \\
&+&  \cos (2\,\alpha )\,\cos (\theta )\,\sin (2\,b)\,
      \sin (2\,a + 2\,\gamma ) dc \nonumber \\
&+&  \cos (2\,b)\,\sin (2\,\alpha )\,\sin (2\,\beta )\,
      \left( 1 - \half\sin^2 (\theta ) \right) dc  \nonumber \\
&-&  \frac{\sqrt{3}}{2}\sin (2\,\alpha )\,\sin (2\,\beta )\,
      \sin^2 (\theta )d\phi \\
\omega^3&=&
d\alpha + \cos (2\,\beta )d\gamma  \nonumber \\
&+&  \cos (2\,\beta )\,
   \left( 1 - \half\sin^2 (\theta ) \right)da   \nonumber \\
&+&  \cos (\theta )\,\sin (2\,\beta )\,
   \sin (2\,a + 2\,\gamma )db  \nonumber \\
&-& \cos (2\,a + 2\,\gamma )\,
        \cos (\theta )\,\sin (2\,b)\,\sin (2\,\beta ) dc \nonumber \\
&+& \cos (2\,b)\,\cos (2\,\beta )\,
      \left( 1 - \half\sin^2 (\theta ) \right) dc \nonumber \\
&-& \frac{\sqrt{3}}{2}\cos (2\,\beta )\,
      \half\sin^2 (\theta )d\phi \\
\omega^4&=&
   \cos (\beta )\,\sin (\alpha  + \gamma )d\theta \nonumber \\
&-&  \half\cos (\beta )\,
      \cos (\alpha  + \gamma )\,\sin (2\,\theta )da  \nonumber \\
&-&  \sin (\beta )\,\sin (2\,a - \alpha  + \gamma )\,
   \sin (\theta )db  \nonumber \\
&+& \cos (2\,a - \alpha  + \gamma )\,
      \sin (2\,b)\,\sin (\beta )\,\sin (\theta ) dc \nonumber \\
&-&  \half\cos (2\,b)\,\cos (\beta )\,\cos (\alpha  + \gamma )\,
         \sin (2\,\theta ) dc \nonumber \\
&-&  \frac{\sqrt{3}}{2}\cos (\beta )\,
      \cos (\alpha  + \gamma )\,\sin (2\,\theta )d\phi \\
\omega^5&=&
\cos (\beta )\,\cos (\alpha  + \gamma )d\theta  \nonumber \\
&+& \half\cos (\beta )\,
      \sin (\alpha  + \gamma )\,\sin (2\,\theta )da  \nonumber \\
&+& 
  \cos (2\,a - \alpha  + \gamma )\,\sin (\beta )\,
   \sin (\theta )db  \nonumber \\
&+& \sin (2\,b)\,\sin (\beta )\,
      \sin (2\,a - \alpha  + \gamma )\,\sin (\theta ) dc \nonumber \\
&+& \half\cos (2\,b)\,\cos (\beta )\,\sin (\alpha  + \gamma )\,
         \sin (2\,\theta ) dc \\
&+& 
  \frac{\sqrt{3}}{2}\cos (\beta )\,
      \sin (\alpha  + \gamma )\,\sin (2\,\theta )d\phi \\
\omega^6&=&
\sin (\beta )\,\sin (\alpha  - \gamma )d\theta  \nonumber \\
&-& 
  \cos (\beta )\,\sin (2\,a + \alpha  + \gamma )\,
   \sin (\theta )db \nonumber \\
&+& \half\cos (\alpha  - \gamma )\,
      \sin (\beta )\,\sin (2\,\theta )da  \nonumber \\
&+& 
  \frac{\sqrt{3}}{2}\cos (\alpha  - \gamma )\,
      \sin (\beta )\,\sin (2\,\theta )d\phi  \nonumber \\
&+&  \cos (\beta )\,
      \cos (2\,a + \alpha  + \gamma )\,\sin (2\,b)\,
      \sin (\theta ) dc \nonumber \\
&+& \half\cos (2\,b)\,
         \cos (\alpha  - \gamma )\,\sin (\beta )\,
         \sin (2\,\theta ) dc \\
\omega^7&=&
-\cos (\alpha  - \gamma )\,\sin (\beta ) d\theta   \nonumber \\
&+&  \half\sin (\beta )\,\sin (\alpha  - \gamma )\,
      \sin (2\,\theta )da  \nonumber \\
&+&  \cos (\beta )\,
   \cos (2\,a + \alpha  + \gamma )\,\sin (\theta )db \nonumber \\
&+&  \cos (\beta )\,\sin (2\,b)\,
      \sin (2\,a + \alpha  + \gamma )\,\sin (\theta ) dc \nonumber \\
&+&  \half\cos (2\,b)\,\sin (\beta )\,\sin (\alpha  - \gamma )\,
         \sin (2\,\theta ) dc \nonumber \\
&+&  \frac{\sqrt{3}}{2}\sin (\beta )\,
      \sin (\alpha  - \gamma )\,\sin (2\,\theta ) d\phi \\
\omega^8 &=&
-\frac{\sqrt{3}}{2}\sin^2 (\theta )da
- \frac{\sqrt{3}}{2}\cos (2\,b)\,
      \sin^2 (\theta )dc
+ \left( 1 - \frac{3}{2}\sin^2 (\theta )
      \right)d\phi
\eea

The right invariant forms are given by:

\bea
\omega^1_r&=&
    - \cos (2\,b)\,\cos (2\,c)\,
        \cos (2\,a + 2\,\gamma )\,\cos (\theta )\,
        \sin (2\,\beta )d\alpha    \nonumber \\
&+&  \cos (\theta )\,\sin (2\,\beta )\,\sin (2\,c)\,
      \sin (2\,a + 2\,\gamma ) d\alpha \nonumber \\
&-&  \cos (2\,\beta )\,\cos (2\,c)\,\sin (2\,b)\,
      \left( 1 - \half\sin^2 (\theta ) \right) d\alpha  \nonumber \\
&-&  \cos (2\,c)\,\sin (2\,b)\,
   \left( 1 - \half\sin^2 (\theta ) \right)d\beta   \nonumber \\
&+&  \cos (2\,a + 2\,\gamma )\,\cos (\theta )\,
      \sin (2\,c) d\gamma \nonumber \\ 
&+&  \cos (2\,b)\,\cos (2\,c)\,\cos (\theta )\,
      \sin (2\,a + 2\,\gamma ) d\gamma    \nonumber \\
&-&  \cos (2\,c)\,\sin (2\,b)\,
   \left( 1 - \half\sin^2 (\theta ) \right) d\theta 
 + \sin (2\,c)db \\
\omega^2_r&=&
  \cos (2\,b)\,\cos (2\,a + 2\,\gamma )\,
      \cos (\theta )\,\sin (2\,\beta )\,\sin (2\,c) d\alpha \nonumber \\
&+& 
     \cos (2\,c)\,\cos (\theta )\,\sin (2\,\beta )\,
      \sin (2\,a + 2\,\gamma ) d\alpha   \nonumber \\
&+&  \cos (2\,\beta )\,\sin (2\,b)\,\sin (2\,c)\,
      \left( 1 - \half\sin^2 (\theta ) \right) d\alpha  \nonumber \\
&+& 
  \sin (2\,b)\,\sin (2\,c)\,
   \left( 1 - \half\sin^2 (\theta ) \right)d\beta    \nonumber \\
&+&
   \left[ \cos (2\,c)\,\cos (2\,a + 2\,\gamma )\,
      \cos (\theta ) - \cos (2\,b)\,\cos (\theta )\,
      \sin (2\,c)\,\sin (2\,a + 2\,\gamma ) \right]d\gamma    \nonumber \\
&+& 
  \sin (2\,b)\,\sin (2\,c)\,
   \left( 1 - \half\sin^2 (\theta ) \right)d\theta + \cos (2\,c)db \\
\omega^3_r&=& - \cos (2\,a + 2\,\gamma )\,
    \cos (\theta )\,\sin (2\,b)\,\sin (2\,\beta ) d\alpha \nonumber \\
&+& \cos (2\,b)\,\cos (2\,\beta )\,
      \left( 1 - \half\sin^2 (\theta ) \right) d\alpha  \nonumber \\
&+& \cos (2\,b)\,
    \left( 1 - \half\sin^2 (\theta ) \right) d\beta \nonumber \\ 
&+& \cos (\theta )\,\sin (2\,b)\,
   \sin (2\,a + 2\,\gamma )d\gamma   \nonumber \\
&+& \cos (2\,b)\,
   \left( 1 - \half\sin^2 (\theta ) \right)d\theta + dc  \\
\omega^4_r&=& \cos (a - c + 2\,\gamma  -3\eta )\,\sin (b)\,
      \sin (2\,\beta )\,\sin (\theta ) d\alpha \nonumber \\ 
&-& \half\cos (b)\,\cos (2\,\beta )\,\cos (a + c +3\eta )\,
         \sin (2\,\theta )d\alpha    \nonumber \\
&-&\half\cos (b)\,\cos (a + c +3\eta )\,
        \sin (2\,\theta ) d\beta \nonumber \\ 
&-& \sin (b)\,\sin (\theta )\,
   \sin (a - c + 2\,\gamma  - 3\eta)d\gamma  \nonumber \\
&-& \half\cos (b)\,\cos (a + c + 3\eta)\,
      \sin (2\,\theta )d\theta 
+ \cos (b)\,\sin (a + c + 3\eta)da \\
\omega^5_r&=& \sin (b)\,\sin (2\,\beta )\,
      \sin (\theta )\,\sin (a - c + 2\,\gamma  - 3\eta) d\alpha \nonumber \\
&+& 
     \half\cos (b)\,\cos (2\,\beta )\,\sin (2\,\theta )\,
         \sin (a + c +3\eta ) d\alpha  \nonumber \\
&+& \half\cos (b)\,\sin (2\,\theta )\,
      \sin (a + c + 3\eta) d\beta \nonumber \\
&+& \cos (a - c + 2\,\gamma  - 3\eta)\,\sin (b)\,
   \sin (\theta )d\gamma   \nonumber \\
&+& 
  \half\cos (b)\,\sin (2\,\theta )\,
      \sin (a + c + 3\eta)d\theta 
+\cos(b)\,\cos(a+c+3\eta)da \\
\omega^6_r&=&  \cos (b)\,\cos (a + c + 2\,\gamma  - 3\eta)\,
      \sin (2\,\beta )\,\sin (\theta ) d\alpha \nonumber \\
&+& 
     \half\cos (2\,\beta )\,\cos (a - c + 3\eta)\,\sin (b)\,
         \sin (2\,\theta ) d\alpha    \nonumber \\
&+& 
  \half\cos (a - c +3\eta )\,\sin (b)\,
      \sin (2\,\theta )d\beta \nonumber \\
&+&  \cos (b)\,\sin (\theta )\,
   \sin (a + c + 2\,\gamma  - 3\eta)d\gamma   \nonumber \\
&+& 
  \half\cos (a - c + 3\eta)\,\sin (b)\,
      \sin (2\,\theta )d\theta 
 - \sin (b)\,\sin (a - c +3\eta )da \\
\omega^7_r&=& \cos (b)\,\sin (2\,\beta )\,
      \sin (\theta )\,\sin (a + c + 2\,\gamma  - 3\eta) d\alpha \nonumber \\
&-&  \half\cos (2\,\beta )\,\sin (b)\,\sin (2\,\theta )\,
         \sin (a - c + 3\eta)  d\alpha  \nonumber \\
&-& 
  \half\sin (b)\,\sin (2\,\theta )\,
      \sin (a - c +3\eta )d\beta  \nonumber \\
&-& \cos (b)\,\cos (a + c + 2\,\gamma  - 3\eta)\,
   \sin (\theta )d\gamma  \nonumber \\
&-& 
  \half\sin (b)\,\sin (2\,\theta )\,
      \sin (a - c +3\eta )d\theta 
 - \cos (a - c + 3\eta)\,\sin (b) da \\
\omega^8_r&=&
 - \frac{\sqrt{3}}{2}\cos (2\,\beta )\,
      \sin^2 (\theta )d\alpha
 - \frac{\sqrt{3}}{2}\sin^2 (\theta ) d\beta
 - \frac{\sqrt{3}}{2}\sin^2 (\theta )d\theta 
 + d\phi
\eea

Thus one may integrate over the whole space or any subspace of the group manifold by the appropriate wedge product of these forms.

\section{Invariant Volume Element}

The group invariant volume element may be calculated in two different ways.  One way is to take an arbitrary $A \in SU(3)$ and find the
 matrix
\begin{equation}
A^{-1}dA
\end{equation}
 of left invariant one-forms and then wedge the 8 linearly independent forms together.  A simpler way to do this is to explicitly compute the wedge product of the invariant one forms calculated above.  This is equivalent to computing the determinant of the matrix of coefficients that appear in (34)-(41).  (the determinant correctly alternates the signs corresponding to the permutations of the differentials in the wedge product.)  The result is
$$
dV = \sin 2\beta \sin 2b \sin 2\theta \sin^2 \theta\; d\alpha\; d\beta\; d\gamma\;
d\theta\; da\; db\; dc\; d\phi.
$$
This agrees with the result of Holland\cite{hol}.  This is determined only up to a
 constant factor since the normalization is determined by setting $\int dV =1$.  It is also easily seen to give left invariant quantities.  For if one constructs this with left invariant vector fields and uses the duality expressed in (24), it is clear that the left hand side (a scalar) is invariant so that the forms must be as well.

The ranges of the angles may be inferred by using two assumptions.  One is that the ranges of the angles in the two $SU(2)$ Euler angle sets are the same as you would expect for the ordinary $SU(2)$ Euler angles.  The other has to do with the volume of the space.  We know that $SU(3)$, as a topological space, is a product of a 3-sphere and a 5-sphere.  There is due to a theorem by H. Hopf that states that compact connected Lie groups has the cohomology (real coefficients) of a product of odd-dimensional of spheres (see for example \cite{Atiyah}).  There are few choices for $SU(3)$. It is semi-simple so $U(1)$ is not one of them.  Its dimensionality is 8, and we can pick out an $SU(2)$, which is a three sphere.  Thus we have only a five sphere remaining.  The volume of a product of two manifolds should be the volume of their product, trivial or not.  (Look at local charts and avoid overlapping.)  The volume of a 5-sphere is $\pi^3$ and of a 3-sphere is $2\pi^2$.  Multiplying these and setting that equal to the integration over the group invariant volume element, and using the first assumption above, one arrives at the following ranges of the angles.
$$
0 \leq \alpha,\gamma,a,c < \pi 
$$
$$
 0 \leq \beta,b,\theta \leq \frac{\pi}{2} \;\;\;\;\;\;\;\;\;\;\; 0 \leq \phi < 2\pi
$$
Therefore any function may be integrated over the group.  

\section*{V  Applications}

The applications are perhaps endless.  Anywhere $SU(3)$ is used, these structures can provide insight.  There are many different $SU(3)$'s in physics, that is, it is used in several different ways.  There is, as stated above, the color gauge group, the approximate flavor symmetry, and the so-called nuclear $SU(3)$.  The latter is based on the shell model and was first introduced by Elliot\cite{Elliot}.  The fact that $SU(3)$ is the symmetry group for the 3-d isotropic harmonic oscillator alone should make for many interesting applications of this material.  With the explicit coordinates and structures derived here (forms and vector fields), one should be able to investigate coset spaces of the group manifold of $SU(3)$ such as $SU(3)/U(1)$ and  $SU(3)/(U(1) \times U(1))$.  These manifolds will be given elsewhere and should prove useful for any 3-d harmonic oscillator potential with the appropriately restricted symmetry.  The manifold $SU(3)/U(1)_Y$, has applications in phenomenological nuclear interactions through the $SU(3)$ Skyrme-Witten model for meson-baryon interactions as well\cite{bnw}.  Here, the coset space is invariant to right actions the $U(1)$ corresponding to the hypercharge($Y$).  This too will be discussed elsewhere.

Given the explicit bases for left and right invariant vector fields and forms, one may construct invariant tensors of arbitrary rank (see for example, \cite{cd}).

Calculations with these forms and vector fields might have seemed quite tedious in the past, given their size.  However, they certainly could be used for computational purposes since they are readily discretized.  Also, they may be manipulated symbolicly using programs such as Mathematica or Maple.  For this purpose, there is a web site under construction with the structures available for download and immediate use on Mathematica(i.e., formatted for Mathematica) at http://www.ph.utexas.edu/\~{ }mbyrd.

\section{Acknowledgments} 

I would, first and foremost, like to thank Prof. L. C. Biedenharn who, as my advisor, first started me on the construction of these structures.  I could not give him too much credit here.  I would also like to thank Prof. E. C. G. Sudarshan for many helpful comments along with Prof. Duane Dicus whose help and support enabled the completion of this paper.  This research was supported in part by the U.S. Department of Energy under Contract No. DE-EG013-93ER40757.

\end{document}